\def\figsize{9.5cm}

\def\rn{}
\def\nn#1 #2{#2. #1}				
\def\nnn#1 #2 #3{#2. #3. #1}			
\def\nnnn#1 #2 #3 #4{#2. #3. #4 #1}		
\def\nnnnn#1 #2 #3 #4 #5{#2. #3. #4 #5. #1}	
\def\dualand{ and\hbox{ }}				
\def\multiand{, and\hbox{ }}				
\def\rf#1;#2;#3;#4;#5 {{\frenchspacing\par\rn#1, #3 {\bf #4}, #5 (#2). \par}}
\def\rrf#1;#2;#3;#4;#5 {{\frenchspacing\rn#1, #3 {\bf #4}, #5 (#2);~}}
\def\rrrf#1;#2;#3;#4;#5 {{\frenchspacing\rn#1, #3 {\bf #4}, #5 (#2).}}
\def\rg#1;#2;#3;#4;#5;#6 {{\frenchspacing\par\rn#1, #3 {\bf #4}, #5 (#2). \par}}
\def\rfbook#1;#2;#3;#4;#5 {{\frenchspacing\par\rn#1, {\it #3} (#5, #4, #2).\par}}
\def\rfprep#1;#2;#3 {{\par\frenchspacing\rn#1, #3 (#2).\par}}
\def\rrfprep#1;#2;#3 {{\frenchspacing\rn#1, #3 (#2);~}}
\def\rrrfprep#1;#2;#3 {{\frenchspacing\rn#1, #3 (#2).}}
\def\rfproc#1;#2;#3;#4;#5;#6 {{\frenchspacing\par\rn#1 #2, in {\it #3}, ed. #4 (#5: #6)\par}}
\def\rfprocp#1;#2;#3;#4;#5;#6;#7 {{\frenchspacing\par\rn#1 #2, in {\it #3}, ed. #4 (#5: #6), p#7\par}}

\def\rg#1;#2;#3;#4;#5;#6 {\par\rn#1 #2, {\it #3}, {\bf #4}, #5 (``#6'') \par}
\def\rf#1;#2;#3;#4;#5 {\par\rn#1, {\it #3}, {\bf #4}, #5 (#2)\par}
\def\rfbook#1;#2;#3;#4;#5 {{\frenchspacing\par\rn#1, {\it #3} (#4: #5, #2)\par}}
\def\rfproc#1;#2;#3;#4;#5;#6 {{\frenchspacing\par\rn#1 #2, in {\it #3}, ed. #4 (#5: #6)\par}}
\def\rfprocp#1;#2;#3;#4;#5;#6;#7 {{\frenchspacing\par\rn#1 #2, in {\it #3}, ed. #4 (#5: #6), p#7\par}}
\def\rfprep#1;#2;#3  {{\par\rn#1, #3, #2\par}}
\def\rfprepp#1;#2;#3 {{\par\rn#1 #2, #3\par}}

\def\etal{{\frenchspacing\it et al.}}

\def\eg{{\frenchspacing\it e.g.}}

\def\bfk{\mbox{\bf k}}
\def\bfr{\mbox{\bf r}}

\def\bfx{\mbox{\bf x}}

\def\bfs{\mbox{\bf s}}

\newcommand{\be}{\begin{equation}}
\newcommand{\ee}{\end{equation}}
\newcommand{\ba}{\begin{eqnarray}}
\newcommand{\ea}{\end{eqnarray}}

\def\ga{\mathrel{\mathpalette\fun >}}
\def\fun#1#2{\lower3.6pt\vbox{\baselineskip0pt\lineskip.9pt
        \ialign{$\mathsurround=0pt#1\hfill##\hfil$\crcr#2\crcr\sim\crcr}}}
        
\def\beq#1{\begin{equation}\label{#1}}
\def\eeq{\end{equation}}
\def\beqa#1{\begin{eqnarray}\label{#1}}
\def\eeqa{\end{eqnarray}}

\def\spose#1{\hbox to 0pt{#1\hss}}
\def\simlt{\mathrel{\spose{\lower 3pt\hbox{$\mathchar"218$}}
     \raise 2.0pt\hbox{$\mathchar"13C$}}}
\def\simgt{\mathrel{\spose{\lower 3pt\hbox{$\mathchar"218$}}
     \raise 2.0pt\hbox{$\mathchar"13E$}}}
\def\simpropto{\mathrel{\spose{\lower 3pt\hbox{$\mathchar"218$}}
     \raise 2.0pt\hbox{$\propto$}}}

\def\ed{\end{document}}

\def\Om{\Omega_m}

\def\beq#1{\begin{equation}\label{#1}}
\def\eeq{\end{equation}}
\def\beqa#1{\begin{eqnarray}\label{#1}}
\def\eeqa{\end{eqnarray}}

\documentclass[twocolumn,amsmath,nofootinbib]{revtex4} 
\usepackage{url}
\begin{document}
\input{epsf.sty}




\def\affilmrk#1{$^{#1}$}
\def\affilmk#1#2{$^{#1}$#2;}


\title{Differentiating dark energy and modified gravity with
galaxy redshift surveys}

\author{Yun Wang}
\address{Homer L. Dodge Department of Physics \& Astronomy, Univ. of Oklahoma, 
440 W.~Brooks St., Norman, OK 73019, USA; wang@nhn.ou.edu}



%
\begin{abstract}
The observed cosmic acceleration today could be due to an
unknown energy component (dark energy),
or a modification to general relativity (modified gravity).
If dark energy models and modified gravity models are
required to predict the same cosmic expansion history $H(z)$,
they will predict {\it different} growth rate for
cosmic large scale structure, $f_g(z)$.
If gravity is not modified, the measured 
$H(z)$ leads to a unique prediction for $f_g(z)$,
$f_g^H(z)$, if dark energy and dark matter are separate. 
Comparing $f_g^H(z)$ with the measured $f_g(z)$
provides a transparent and straightforward test of gravity.  
We show that a simple $\chi^2$ test provides a general
figure-of-merit for our ability to distinguish between
dark energy and modified gravity given the measured $H(z)$
and $f_g(z)$.
We find that a magnitude-limited NIR
galaxy redshift survey covering $>$10,000 (deg)$^2$ 
and the redshift range of $0.5<z<2$ can be used
to measure $H(z)$ to 1-2\% accuracy via baryon acoustic oscillation 
measurements, and $f_g(z)$ to the accuracy of a few percent 
via the measurement of redshift-space distortions 
and the bias factor which describes how
light traces mass.
We show that if the $H(z)$ data are fit by both a DGP gravity model and
an equivalent dark energy model that predict the same $H(z)$, 
a survey area of 11,931$\,$(deg)$^2$ is required
to rule out the DGP gravity model at the 99.99\% confidence level.
It is feasible for such a galaxy redshift survey to be carried out by
the next generation space missions from NASA and ESA,
and it will revolutionize our understanding of the universe
by differentiating between dark energy and modified gravity.

\end{abstract}


  
\maketitle



\section{Introduction}
The observed cosmic acceleration today \cite{Riess98,Perl99}
could be due to an unknown energy component (dark energy, 
\eg, \cite{quintessence}), or a modification to general relativity 
(modified gravity, \eg, \cite{modifiedgravity,DGPmodel}).
Ref.\cite{reviews} contains reviews with more complete lists of references.
Illuminating the nature of dark energy is one of the most exciting 
challenges in cosmology today.

The cosmic expansion history, $H(z)=(d\ln a/dt)$ ($a$ is the cosmic 
scale factor), and the growth rate for cosmic large scale structure, 
$f_g(z)=d\ln \delta/d\ln a$ 
[$\delta=(\rho_m-\overline{\rho_m})/\overline{\rho_m}$],
are two functions of redshift $z$ 
that can be measured from cosmological data.
They provide independent and complementary probes
of the nature of the observed cosmic acceleration \cite{DGP,otherfg}.
The precisely measured $H(z)$ and $\Omega_m$ lead to a unique prediction 
for $f_g(z)$ in the absence of modified gravity,
$f_g^H(z)$, if dark energy and dark matter are separate.
Comparing $f_g^H(z)$ with the measured $f_g(z)^{obs}$
provides a transparent and straightforward test of gravity (see Fig.1).
If gravity is not modified, $H(z)$ and $f_g(z)$
together provide stronger constraints on dark energy
models \cite{fg_use}. 

Using the VVDS data,
Ref.\cite{Guzzo07} demonstrated that a magnitude-limited
galaxy redshift survey can be used to measure $f_g(z)$
via measurements of redshift-space distortion parameter
\be
\beta(z)=\frac{f_g(z)}{b(z)}
\ee
and the bias parameter
$b(z)$ (which describes how light traces mass)
from galaxy clustering.
In this paper we show that a feasible, sufficiently wide and deep
magnitude-limited galaxy redshift survey will allow us to unambiguously 
differentiate between dark energy and modified gravity by providing
precise measurements of $H(z)$ and $f_g(z)$ (see Fig.1).

\begin{figure} 
\vskip-1.3cm
\centerline{\epsfxsize=\figsize\epsffile{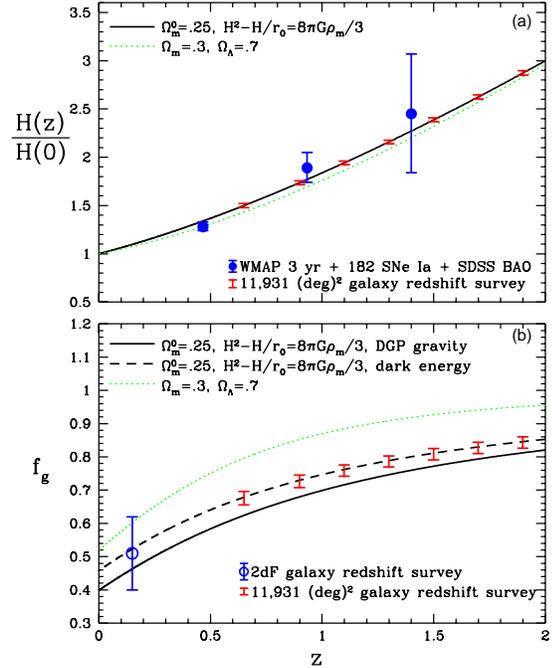}}
\vskip-0.3cm
\caption[1]{\label{Hzfg}\footnotesize%
Current and expected future measurements of the 
cosmic expansion history $H(z)=H_0 E(z)$ and the growth rate of 
cosmic large scale structure 
$f_g(z)=d\ln \delta/d\ln a$ ($\delta=(\rho_m-\overline{\rho_m})/
\overline{\rho_m})$, $a$ is the cosmic scale factor).
Note that the fiducial model assumed for the future galaxy redshift
survey is a dark energy model with the same $H(z)$ as that of
the DGP model. These two models have identical expansion histories
$H(z)$ [solid line in panel (a)], but very different growth
rates $f_g(z)$ [solid and dashed lines in panel (b)].
}
\end{figure}

\section{Models}
If the present cosmic acceleration is caused by dark energy,
$E(z) \equiv H(z)/H_0=[\Omega_m (1+z)^3 + \Omega_k (1+z)^2 +\Omega_X X(z)
]^{1/2}$, where $X(z)\equiv \rho_X(z)/\rho_X(0)$, with $\rho_X(z)$
denoting the dark energy density.
The linear growth rate  $f_g\equiv d\ln D_1/d\ln a$ 
is determined by solving the equation for $D_1=\delta^{(1)}(\bfx,t)/\delta(\bfx)$, 
\be
\label{eq:fg}
D_1''(\tau) + 2E(z)D_1'(\tau) - {3\over 2}\Om (1+z)^3D_1 = 0,
\ee
where primes denote $d/d(H_0 t)$, and we have assumed that dark
energy and dark matter are separate.

In the simplest alternatives to dark energy,
the present cosmic acceleration is caused by a modification
to general relativity. The only rigorously worked example
is the DGP gravity model \cite{DGPmodel,DGP},
which can be described by a modified 
Friedmann equation:
\be
H^2 - \frac{H}{r_0}=\frac{8\pi G \rho_m}{3},
\label{eq:H(z)_DGP}
\ee
where $r_0$ is a parameter in DGP gravity, and 
$\rho_m(z)=\rho_m(0)(1+z)^3$. Solving Eq.(\ref{eq:H(z)_DGP}) 
gives 
\be
E(z)=\frac{H(z)}{H_0}=\frac{1}{2} \left\{ 
\frac{1}{H_0 r_0}+\left[ \frac{1}{(H_0 r_0)^{2}} 
+ 4 \Omega_m^0 (1+z)^3 
\right]^{1/2} \right\},
\ee
with $\Omega_m^0 \equiv \rho_m(0)/\rho_c^0$, $\rho_c^0 \equiv
3H_0^2/(8\pi G)$. The added superscript ``0'' in
$\Omega_m^0$ denotes that this is the matter fraction today
in the DGP gravity model.
Note that consistency at $z=0$, $E(0)=1$ 
requires that 
\be
H_0 r_0 = \frac{1}{1-\Omega_m^0}, 
\ee
so the DGP gravity model is parametrized by a single parameter,
$\Omega_m^0$.
The linear growth factor in the DGP gravity model is
given by \cite{DGP}
\be
\label{eq:fg_DGP}
D_1''(\tau) + 2E(z)D_1'(\tau) - {3\over 2}\Om (1+z)^3D_1 
\left(1+ \frac{1}{3\alpha_{DGP}} \right)= 0,
\ee
where 
\be
\alpha_{DGP}=\frac{1-2H_0r_0+ 2(H_0 r_0)^2}{1-2H_0r_0}.
\ee

The dark energy model equivalent of the DGP gravity model
is specified by requiring 
\be
\frac{8\pi G\rho_{de}^{eff}}{3}=\frac{H}{r_0}.
\ee
Eq.(\ref{eq:H(z)_DGP}) and the conservation of energy and
momentum equation, 
\be
\dot{\rho}_{de}^{eff}+3(\rho_{de}^{eff}
+p_{de}^{eff})H=0,
\ee
imply that \cite{DGP}
\be
w_{de}^{eff}=-\frac{1}{1+\Omega_m(a)},
\ee
where
\be
\Omega_m(a)  \equiv \frac{8\pi G \rho_m(z)}{3 H^2}
= \frac{\Omega_m^0 (1+z)^3}{E^2(z)}.
\ee
As $a\rightarrow 0$, $\Omega_m(a) \rightarrow 1$,
and $w_{de}^{eff} \rightarrow -0.5$.
As $a \rightarrow 1$, $\Omega_m(a) \rightarrow \Omega_m^0$,
and $w_{de}^{eff} \rightarrow -1/(1+\Omega_m^0)$.
This means that the matter transfer function for
the dark energy model equivalent of viable DGP gravity model
($\Omega_m^0<0.3$ and $w\leq -0.5$) is very close
to that of the $\Lambda$CDM model 
at $k\ga 0.001\,h\,$Mpc$^{-1}$.\cite{Ma99}

It is very easy and straightforward to integrate 
Eqs.(\ref{eq:fg}) and (\ref{eq:fg_DGP}) to obtain
$f_g$ for dark energy models and DGP gravity models,
with the initial condition that for $a\rightarrow 0$, $D_1(a) = a$
(which assumes that the dark energy or modified gravity
are negligible at sufficiently early times).
There are well known approximations to $f_g$,
with $f_g(z) = \Omega_m(a)^{6/11}$ for dark energy
models \cite{WangStein},
and $f_g(z) = \Omega_m(a)^{2/3}$ for DGP gravity models
\cite{DGP}.
Fig.2 shows that these 
powerlaw approximations of $f_g$
are not sufficiently accurate for future galaxy redshift
surveys that can measure $f_g$ to a few percent accuracy
in $\Delta z=0.2$ redshift bins.

\begin{figure} 
\vskip 0.1cm
\centerline{\epsfxsize=\figsize\epsffile{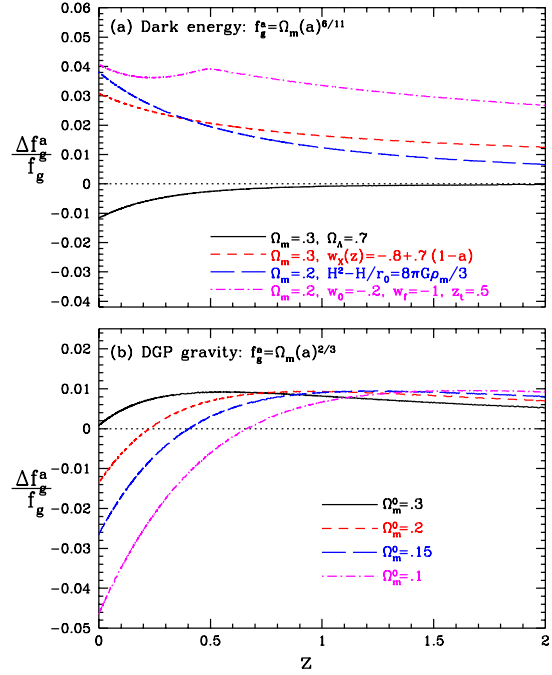}}
\caption[1]{\label{fgOa}\footnotesize%
The accuracy of approximate expressions for $f_g(z)$
for various models.
}
\end{figure}

\section{Analysis Technique}

Galaxy redshift surveys allow us to measure both $H(z)$ and $f_g(z)$
through baryon acoustic oscillation (BAO) measurements 
\cite{BG03,SE03,Eisen05,BAO} and redshift-space distortion measurements \cite{Guzzo07}.
BAO in the observed galaxy power
spectrum have the characteristic scale determined by the
comoving sound horizon at recombination, which is
precisely measured by the cosmic microwave background (CMB) 
anisotropy data \cite{Spergel06}.
Comparing the observed BAO scales with the expected values
gives $H(z)$ in the radial direction, and $D_A(z)$ [the angular
diameter distance $D_A(z)=r(z)/(1+z)$, where
$r(z)$ is the coordinate or comoving distance] in the transverse direction.
We will only estimate the accuracy to which $H(z)$
and $f_g(z)$ can be determined from galaxy redshift surveys
in dark energy models (the error bars in Fig.1).\footnote{\cite{Roman}
gives a more precise treatment of redshift-space distortions,
and \cite{Stab} studies power spectra in alternative gravity models.}

The observed power spectrum is reconstructed using a particular reference 
cosmology, including the effects of bias and redshift-space distortions \cite{SE03}:
\ba
\label{eq:P(k)}
P_{obs}(k^{ref}_{\perp},k^{ref}_{\parallel}) &=&
\frac{\left[D_A(z)^{ref}\right]^2  H(z)}{\left[D_A(z)\right]^2 H(z)^{ref}}
\, b^2 \left( 1+\beta\, \mu^2 \right)^2
\cdot \nonumber\\
& \cdot& \left[ \frac{G(z)}{G(0)}\right]^2 P_{matter}(k)_{z=0}+ P_{shot},
\ea
where the growth factor $G(z)$ and the growth rate
$f_g(z)=\beta b(z)$ are related via
$f_g(z)=d\ln G(z)/d\ln a$, and 
$\mu = \bfk \cdot \hat{\bfr}/k$, with $\hat{\bfr}$ denoting the unit
vector along the line of sight; $\bfk$ is the wavevector with $|\bfk|=k$.
Hence $\mu^2=k^2_{\parallel}/k^2=k^2_{\parallel}/(k^2_{\perp}+k^2_{\parallel})$.
The values in the reference cosmology are denoted by the subscript ``ref'',
while those in the true cosmology have no subscript.
Note that 
\be
k^{ref}_{\perp}=k_\perp\,\frac{ D_A(z)}{D_A(z)^{ref}}, \hskip 0.5cm
k^{ref}_{\parallel}=k_\parallel\,\frac{H(z)^{ref}}{H(z)}.
\ee
Eq.(\ref{eq:P(k)}) characterizes the dependence of the observed galaxy power
spectrum on $H(z)$ and $D_A(z)$ due to BAO, as well as 
the sensitivity of a galaxy redshift survey to the redshift-space 
distortion parameter $\beta$ \cite{Kaiser}.

To study the expected impact of future galaxy redshift surveys,
we use the Fisher matrix formalism.
In the limit where the length scale corresponding to
the survey volume is much larger than
the scale of any features in $P(k)$, 
we can assume that the likelihood function for the band powers of a 
galaxy redshift survey is Gaussian \cite{FKP}. 
Then the Fisher matrix can be approximated as \citep{Tegmark97}
\be
F_{ij}= \int_{k_{min}}^{k_{max}}
\frac{\partial\ln P(\bfk)}{\partial p_i}
\frac{\partial\ln P(\bfk)}{\partial p_j}\,
V_{eff}(\bfk)\, \frac{d \bfk^3}{2\, (2\pi)^3}
\label{eq:full Fisher}
\ee
where $p_i$ are the parameters to be estimated from data, and 
the derivatives are evaluated at parameter values of the
fiducial model. The effective volume of the survey
\ba
V_{eff}(k,\mu) &=&\int d\bfr^3 \left[ \frac{n(\bfr) P(k,\mu)}{ n(\bfr) P(k,\mu)+1}
\right]^2
\nonumber\\
&=&\left[ \frac{ n P(k,\mu)}{n P(k,\mu)+1} \right]^2 V_{survey},
\ea
where the comoving number density $n$ is assumed to only depend on
the redshift for simplicity.
Note that the Fisher matrix $F_{ij}$ is the inverse of the covariance matrix
of the parameters $p_i$ if the $p_i$ are Gaussian distributed.
Eq.(\ref{eq:full Fisher}) propagates the measurement
error in $\ln P(\bfk)$ (which is proportional to $[V_{eff}(\bfk)]^{-1/2}$)
into measurement errors for the parameters $p_i$.

Since we do not include nonlinear effects, we only consider
wavenumbers smaller than a minimum value of non-linearity.
Following \cite{BG03}, we take $k_{min}=0$, and $k_{max}$ given by
requiring that the variance of matter fluctuations
in a sphere of radius $R$, $\sigma^2(R)= 0.35$, for $R=\pi/(2k_{max})$.
We will also give results for $\sigma^2(R)= 0.2$ for comparison.
In addition, we impose a uniform upper limit of $k_{max}\leq 0.2\,h$Mpc$^{-1}$, 
to ensure that we are only considering the conservative linear regime
essentially unaffected by nonlinear effects.
\cite{nonlinear1} shows that nonlinear effects can be accurately
taken into account. \cite{nonlinear2} shows that
the BAO signal is {\it boosted} when these effects are properly
included in the Hubble Volume simulation.
We assume $\Omega_b=0.045$, $h=0.7$, $b=1$,
and $nP=3$ \cite{BG03};
this is conservative since $nP>3$ at any redshift
for a magnitude-limited survey.

The observed galaxy power spectrum in a given redshift shell centered
at redshift $z_i$ can be described by a set of parameters, 
\{$H(z_i)$, $D_A(z_i)$, $\overline{G(z_i)}$, $\beta(z_i)$, $P_{shot}^i$, $n_S$, $\omega_m$,
$\omega_b$\}, where $n_S$ is the power-law index of the primordial
matter power spectrum, $\omega_m=\Omega_m h^2$, and $\omega_b=\Omega_b h^2$
($h$ is the dimensionless Hubble constant). 
Note that $P(k)$ does {\it not} depend on $h$ if $k$ is in units of Mpc$^{-1}$, 
since the matter transfer function $T(k)$ only depends on $\omega_m$ and
$\omega_b$ \cite{EisenHu98},\footnote{Massive neutrinos can suppress
the galaxy power spectrum amplitudes by $\ga 4$\% on BAO
scales \cite{neutrino}. It will be important for future
work to quantify the effect of massive neutrinos
on the measurement of $H(z)$ and $f_g(z)$.}  
if the dark energy dependence of $T(k)$ can be neglected. 
Since $G(z)$, $b$, and the power spectrum normalization $P_0$ are 
completely degenerate in Eq.({\ref{eq:P(k)}}), we have defined
$\overline{G(z_i)}\equiv b\,G(z)\,P_0^{1/2}/G(0)$.

The square roots of diagonal elements of the inverse of the full Fisher matrix of 
Eq.(\ref{eq:full Fisher}) gives the estimated smallest possible measurement
errors on the assumed parameters. 
The parameters of interest are \{$H(z_i)$, $D_A(z_i)$, $\beta(z_i)$\},
all other parameters are marginalized over.
Note that the estimated errors we obtain here are {\it independent}
of cosmological priors,\footnote{Priors on $\omega_m$, 
$\omega_b$, $\Omega_k$, and $n_S$ will be required to obtain the 
errors on dark energy parameters.}
thus scale with (area)$^{-1/2}$ for a fixed survey depth.

Fig.1 shows the errors on $H(z)$ and $f_g(z)=\beta(z) b(z)$ 
for a dark energy model that gives the same $H(z)$ as a
DGP gravity model with the same $\Omega_m^0$,
for a redshift survey covering 11,931 (deg)$^2$, and 
the redshift range $0.5<z<2$ [$\sigma^2(R)= 0.35$ assumed]. 
Note that the $D_A(z)$ measured from the same redshift survey provides
additional constraints on $H(z)$ that can be used for cross-checking
to eliminate systematic effects.
We have neglected the very weak dependence of the transfer function
on dark energy at very large scales in this model \cite{Ma99}, and
added $\Delta\ln b=0.01\,\{$(area)/[28,600$\,$(deg)$^2$]$\}^{-1/2}$
in quadrature to the estimated error on $\beta$.\footnote{This 
$\Delta\ln b$ estimate comes from extrapolating 2dF measurement 
of $b=1.04\pm0.11$ at $z\sim 0.15$ for an effective survey area of 
1300$\times$127000/245591=672 (deg)$^2$ \cite{2dFbias},
and assuming a factor of 1.6 improvement for a NIR space mission
that can detect galaxies at a much higher number density. 
This $\Delta\ln b$ estimate is comparable (and larger) than 
that estimated by \cite{Dolney06} for imaging surveys at $z<2$.}

\cite{bias} developed the method for measuring $b(z)$
from the galaxy bispectrum, which was applied by \cite{2dFbias}
to the 2dF data.
Assuming that \cite{Fry93}
\be
\delta_g= b \delta(\bfx)+ \frac{1}{2}\,b_2 \delta^2(\bfx),
\ee
the galaxy bispectrum
\ba
\langle \delta_{g\bfk_1} \delta_{g\bfk_2} \delta_{g\bfk_1}\rangle
&=&(2\pi)^3 \left\{P_g(\bfk_1) P_g(\bfk_2)
\left[\frac{J(\bfk_1,\bfk_2)}{b}
\right.\right.\nonumber\\
&& \left.\left.+\frac{b_2}{b^2}\right]+cyc.\right\} \delta^D(\bfk_1+\bfk_2+\bfk_3),
\nonumber\\
\ea
where $J$ is a function that depends on the shape of the
triangle formed by ($\bfk_1$, $\bfk_2$, $\bfk_3$) in
$\bfk$ space, but only depends very weakly on cosmology \cite{bias}.

Ref.\cite{BG03} used Monte Carlo N-body simulation to study the
extraction of the BAO scales.
For comparison, we calculated \{$H(z_i)$, $D_A(z_i)$\} for the
same fiducial model as considered by \cite{BG03} (with the
same assumptions and cutoffs in $k$),
and obtained results that are within 30\% of the values
given by the fitting formulae from \cite{Blake06}.
This is reassuring, as it validates the approach of using the Fisher
matrix formalism to forecast the parameter accuracies for
future redshift surveys. \footnote{Ref.\cite{SE05} found similar
agreement in their comparison.}

\section{Observational methods}
$H(z)$ can be probed using multiple techniques.
It can be measured using Type Ia supernovae (SNe Ia)
as cosmological standard candles \cite{SNe Ia}.
CMB and large scale structure data provide constraints on cosmological
parameters that help tighten the constraints on $H(z)$ \cite{WangPia07}.
Fig.1(a) shows the $H(z)$ given by Eq.(\ref{eq:H(z)_DGP}) with
$\Omega_m^0=0.25$ (solid line), as well as a cosmological constant model
with $\Omega_m=0.3$, $\Omega_{\Lambda}=0.7$ (dotted line).
Clearly, both these fit the constraints on $H(z)$ from
current data \cite{WangPia07} (no priors assumed).\footnote{Ref.\cite{WangPia07} 
uses WMAP three year data \cite{Spergel06}, 182 type Ia supernovae \cite{SNeIa07},
and the SDSS baryon acoustic scale measurement \cite{Eisen05}.}

BAO measurements from a very wide and deep galaxy redshift
survey provide a direct precise measurement of $H(z)$
[see Fig.1(a)]. Suppose $H(z)$ is measured to be
$H^2-H/r_0=8\pi G\rho_m/3$ [see solid
line in Fig.1(a)] and $\Omega_m$ is known accurately, 
Eq.(\ref{eq:fg}) yields a unique prediction for $f_g(z)$, 
$f_g^{H}(z)$, assuming that gravity is {\it not} 
modified [see the dashed line in Fig.1(b)].

The measurement of $f_g(z)$ can be obtained through
independent measurements of 
$\beta=f_g(z)/b$ and $b(z)$ \cite{Guzzo07}.
The parameter $\beta$ 
can be measured directly from galaxy redshift survey data
by studying the observed redshift-space correlation function
\cite{2dFbeta,beta}. 
The bias factor $b(z)$ can be measured
by studying galaxy clustering properties (for example,
the galaxy bispectrum) from
the galaxy redshift survey data \cite{2dFbias}.
Independent measurements of $\beta(z)$ and $b(z)$
have only been published for the 2dF data 
\cite{2dFbeta,2dFbias,NP07}.

Fig.1(b) shows the $f_g(z)$ for the DGP gravity model
with $\Omega_m^0=0.25$ (solid line), as well as a dark energy model
that gives the same $H(z)$ for the same $\Omega_m^0$ (dashed line).
The cosmological constant model from Fig.1(a) 
is also shown (dotted line).
Clearly, current data can not differentiate between dark energy
and modified gravity.

A very wide and deep galaxy redshift survey provides
measurement of $f_g(z)$ accurate to a few percent
[see Fig.1(b)]; this will allow an unambiguous distinction
between dark energy models and modified gravity models
that give identical $H(z)$ [see the solid and dashed lines
in Fig.1(b)]. A simple $\chi^2$ test can provide a general
figure-of-merit for our ability to distinguish between
dark energy and modified gravity models that fit the measured $H(z)$
but predict different $f_g(z)$. If the measurement errors 
are normally distributed,
$\Delta \chi^2 \equiv \chi^2(\bfs)-\chi^2(\bfs_0)$ is
distributed as a chi-square distribution with $n$ degrees
of freedom ($n$ is the number of data points), where 
$\bfs$ is the test model, and $\bfs_0$ is the bestfit
model measured from data.
$P(\chi^2|n)=99.99$\% corresponds to $\Delta \chi^2=29.877$ for $n=7$.
Assuming that $\chi^2(\bfs_0)=n$, we find that $\chi^2(\bfs)=36.877$.
In Fig.1, we assume that the true model is a dark energy model with
$\Omega_m^0=0.25$, $H^2-H/r_0=8\pi G\rho_m/3$, with
$Hr_0=1/(1-\Omega_m^0)$.
For a linear cutoff given by $\sigma^2(R)= 0.35$ (or 0.2),
a survey covering 11,931$\,$(deg)$^2$ would rule out the
DGP gravity model that gives the same
$H(z)$ and $\Omega_m^0$ at 99.99\% (or 95\%) C.L.;
a survey covering 13,912$\,$(deg)$^2$ would rule out the
DGP gravity model at 99.999\% (or 99\%) C.L..

\section{Conclusions}

Discovering the nature of dark energy has been identified
as a high priority by both NASA and ESA.
A magnitude-limited NIR galaxy redshift survey, covering 
$>$10,000 (deg)$^2$ and the redshift range $0.5<z<2$,
can be feasibly carried out by a space mission 
that uses MEMS technology to obtain 5000-10,000 galaxy
spectra simultaneously \cite{JEDI,SPACE}.
The low background from space enables very short exposure times
to obtain galaxy spectra to $z\sim 2$,
making it practical to carry out a
magnitude-limited NIR galaxy redshift survey 
over $>$10,000 (deg)$^2$ in only a few years. 
A magnitude-limited galaxy redshift survey 
over $>10,000$ (deg)$^2$ will enable
robust and precise determination of $b(z)$ 
using multiple techniques 
and with sufficient statistics \cite{bias,2dFbias,Guzzo07}.
This is critical for determining $f_g(z)$ using 
measurements of redshift-space distortions.
Such a survey will also enable rigorous study
of the systematic uncertainties of BAO, and accurate
measurements of redshift-space distortions.

Ref.\cite{Knox06} studied the use of weak lensing
shear maps to differentiate between dark energy 
and modified gravity, complementary to
what we have studied in this paper.
While both weak lensing surveys
and galaxy redshift surveys can provide accurate
measurements of $H(z)$ (if the systematic uncertainties
are properly modeled and controlled), galaxy redshift surveys
can potentially provide the most accurate measurement
of $f_g(z)$ [compare Fig.2 of \cite{Knox06} with
Fig.1 of this paper, noting that $f_g(z)=d\ln G(z)/d\ln a$].

We have shown that a magnitude-limited 
NIR galaxy redshift survey covering $>$10,000 (deg)$^2$
and $0.5<z<2$ can provide precise measurements 
of the cosmic expansion history $H(z)$, and the growth 
rate of cosmic large scale structure $f_g(z)$. 
These provide {\it model-independent} constraints
on dark energy 
and the nature of gravity.
The precisely measured $H(z)$ can be used to
predict $f_g(z)$ expected in the absence of modified
gravity, $f_g^H(z)$, if dark energy and dark matter
are separate.
Comparing $f_g^H(z)$ with $f_g(z)^{obs}$ provides a 
transparent and powerful probe of modified gravity.
This will allow us to illuminate
the nature of the observed cosmic acceleration
by differentiating between dark energy and modified
gravity [see Fig.1]. 
A magnitude-limited survey covering 11,931$\,$(deg)$^2$ can
rule out the DGP gravity model at the 99.99\% 
confidence level.\footnote{Such a survey would allow us to 
distinguish between dark energy and modified
gravity even if dark energy is clustered such that
$f_g$, bias, and redshift distortions are scale-dependent \cite{DEcluster}, 
since a dark energy 
model and a modified gravity model generally have
different redshift dependences of the
modified growth rate,
and the data of such a survey can be analyzed 
in multi redshift slices, on multi scales, and using
different populations of galaxies.}
If this technologically feasible 
survey is carried out by a space mission, it will
have a revolutionary effect on our understanding of
the universe.

{\bf Acknowledgements:}
I thank Gigi Guzzo for helpful comments on a draft of this paper,
and Chris Blake, Craig Wheeler, and Eiichiro Komatsu
for useful discussions.


\begin{thebibliography}{99}

\bibitem{Riess98}
\rf\nnn Riess A G {\etal};1998;Astron. J.;116;1009

\bibitem{Perl99} 
\rf\nn Perlmutter S {\etal};1999;ApJ;517;565

\bibitem{quintessence}
\rrf\nn Freese K {\etal};1987;Nucl.Phys.;B287;797
\rrf\nnnn Peebles P J E\dualand\nn Ratra B;1988;ApJ;325;L17
\rrf\nn Wetterich C;1988;Nucl.Phys.;B302;668 
\rrf\nnn Frieman J A {\etal};1995;PRL;75;2077 
\rrf\nn Caldwell R, \nn Dave R\multiand\nnn Steinhardt P J;1998;PRL;80;1582


\bibitem{modifiedgravity}  
\rrf\nn Sahni V \dualand\nn Habib S;1998;PRL;81;1766
\rrf\nn Parker L\dualand\nn Raval A;1999;PRD;60;063512
\rrf\nn Deffayet C;2001;Phys.Lett.B;502;199 
J-P. Uzan, F. Bernardeau, Phys. Rev. D64 (2001) 083004;	
\rrf\nn Freese K\dualand\nn Lewis M;2002;Phys.Lett.B;540;1 
\rrrf\nnn Onemli V K  \dualand\nnn Woodard R P;2004;PRD;70;107301


\bibitem{DGPmodel} 
\rf\nn Dvali G, \nn Gabadadze G, \nn Porrati M;2000;PLB;485;208


\bibitem{reviews}
\rrf\nn Padmanabhan T;2003;Phys.Rep.;380;235 
\rrf\nnnn Peebles P J E\dualand\nn Ratra B;2003;Rev.Mod.Phys.;75;55
\rrf\nn Sahni V \dualand\nn Starobinsky A;2006;IJMPD;15;2105
\rrf\nnn Copeland E J, \nn Sami M\multiand\nn Tsujikawa S;2006;IJMPD;15;1753
\rrf\nn Ruiz-Lapuente P;2007;Class. Quantum. Grav.;24;R91 
\rrrfprep\nn Ratra B \dualand\nnn Vogeley M S;2007;arXiv:0706.1565
V. Sahni, A. Starobinsky, IJMPD 15 (2006) 2105 

\bibitem{DGP}
\rrf\nn Lue A, \nn Scoccimarro R\multiand\nnn Starkman G D;2004;PRD;69;124015
\rrrf\nn Lue A;2006;Physics Report;423;1


\bibitem{Knox06}
\rf\nn Knox L, \nnn Song Y S, \multiand\nnn Tyson J A;2006;PRD;74;023512

\bibitem{otherfg}
A. F. Heavens, T.D. Kitching, L. Verde, MNRAS, 380, 1029 (2007);
P. Zhang, M. Liguori, R. Bean, and S. Dodelson,
Phys.Rev.Lett. 99 (2007) 141302;
D. Sapone, L. Amendola, arXiv:0709.2792 [ps, pdf, other]


\bibitem{fg_use}
\rrf\nnn Knop R A {\etal};2003;ApJ;598;102
\rrf\nn Wang Y\dualand\nn Mukherjee P;2004;ApJ;606;654
\rrrf\nn Wang Y\dualand\nn Tegmark T;2004;PRL;92;241302

\bibitem{Guzzo07}
L. Guzzo, {\etal}, {\it Nature}, 451, 541 (2008)



\bibitem{Ma99}
\rf\nnn Ma C P, \nnn Caldwell R R, \nn Bode P\multiand\nn Wang L;
1999;ApJ;521;L1

\bibitem{WangStein}
\rf\nn Wang L\dualand\nnn Steinhardt P J;1998;ApJ;508;483

\bibitem{Spergel06}
\rf\nnn Spergel D N {\etal};2007;ApJS;170;377

\bibitem{BG03}
\rf\nn Blake C\dualand\nn Glazebrook K;2003;ApJ;594;665

\bibitem{SE03}
\rf\nn Seo H\dualand\nnn Eisenstein D J;2003;ApJ;598;720


\bibitem{BAO}
See for example,
\rrf\nn White M;2005;Astropart. Phys.;24;334
\rrf\nn Huetsi G;2006;A\&A;449;891
\rrf\nn Wang Y;2006;ApJ;647;1  
\rrrfprep\nn Angulo R {\etal};2007;astro-ph/0702543

\bibitem{Roman}
R. Scoccimarro, PRD, {\bf 70}, 083007 (2004)

\bibitem{Stab}
\rf\nnn	Stabenau H F\dualand\nn Jain B;
2006;PRD;74;084007

\bibitem{SE05}
\rf\nn Seo H\dualand\nnn Eisenstein D J;2005;ApJ;633;575

\bibitem{Kaiser}
\rf\nn Kaiser N;1987;MNRAS;227;1


\bibitem{FKP}
\rf\nnn Feldman H A, \nn Kaiser N\multiand\nnn Peacock J A;1994;ApJ;426;23

\bibitem{Tegmark97}
\rf\nn Tegmark M;1997;PRL;79;3806

\bibitem{neutrino}
\rrf\nnn Eisenstein D J\dualand\nn Hu W;
1999;ApJ;511;5
\rrrf\nn Hu W, \nnn Eisenstein D J\multiand\nn Tegmark M;
1998;PRL;80;5255

\bibitem{nonlinear1}
\rrf\nn Jeong D\dualand\nn Komatsu E;
2006;ApJ;651;619
M. Crocce, and R. Scoccimarro, arXiv:0704.2783;
R. E. Smith, R. Scoccimarro, and R. K. Sheth, astro-ph/0703620;


\bibitem{nonlinear2}
\rf\nnn	Koehler R S, \nn Schuecker P\multiand\nn Gebhardt K;2007;A\&A;462;7
	
\bibitem{EisenHu98}
\rf\nn Eisenstein D\dualand\nn Hu W;1998;ApJ;496;605

\bibitem{Blake06}
\rf\nn Blake C {\etal};2006;MNRAS;365;255



\bibitem{SNe Ia}
\rrf\nnn Phillips M M; 1993;ApJ;413;L105 

\bibitem{WangPia07}
Y. Wang, and P. Mukherjee, Phys. Rev. D 76, 103533 (2007)

\bibitem{SNeIa07}
\rrf\nn Astier P {\etal};2006;Astron. Astrophys.;447;31
\rrrf\nnn Riess A G {\etal};2007;ApJ;659;98	

\bibitem{Eisen05}
\rrf\nn Eisenstein D {\etal};2005;ApJ;633;560

\bibitem{2dFbeta}
\rf\nn Hawkins E {\etal};2003;MNRAS;346;78

\bibitem{beta}
\rrf\nn Tegmark M {\etal};2006;PRD;74;123507
\rrf\nnn Ross N P {\etal};2007;MNRAS;381;573
J. da Angela {\etal}, astro-ph/0612401.

\bibitem{Fry93}
\rf\nnn Fry J N\dualand\nn Gaztanaga E;
1993;ApJ;413;447

\bibitem{bias}
\rf\nn Matarrese S, \nn Verde L\multiand\nnn Heavens A F;
1997;MNRAS;290;651

\bibitem{2dFbias}
L. Verde, {\etal}, MNRAS, 335, 432 (2002);
http://www.mso.anu.edu.au/2dFGRS/	

\bibitem{Dolney06}
\rf\nn Dolney D, \nn Jain B\multiand\nn Takada M;2006;MNRAS;366;884
	
\bibitem{NP07}
S. Nesseris, \& L. Perivolaropoulos, arXiv:0710.1092 


\bibitem{JEDI}
\rrf\nn Wang Y {\etal};2004;BAAS;v36;n5;1560
\rrfprep\nn Crotts A {\etal};2005;astro-ph/0507043
\rrrf\nn Cheng E\dualand\nn Wang Y {\etal};2006;Proc. of SPIE;Vol. 6265;626529 

\bibitem{SPACE}
M. Robberto, A. Cimatti, and the SPACE science team,
Venice 2007 Conf. Proc., to appear on Il Nuovo Cimento,
arXiv:0710.3970

\bibitem{DEcluster}
\rrf\nn Hu W;2002;PRD;65;023003
\rrf\nn Gordon C, \dualand\nn Hu W;2004;PRD;70;083003
L. Hui, and K.P. Parfrey, arXiv:0712.1162

	
\end{thebibliography}
\end{document}